\documentclass[12pt]{article}
\usepackage{amssymb,amsmath,euscript,latexsym,amsfonts,epsfig}

\newcommand{\vect}[1]{{\boldsymbol #1}}

\renewcommand{\P}{{\cal P}}

\begin{document}
\title{
\textsc{\bf Numerical Experiment on Interference for Macroscopic Particles}
$~$\\}
\author{
\textsf{A.Yu.~Khrennikov}
\\
International Center for Mathematical Modeling\\
in Physics, Engineering and Cognitive science\\
MSI, V\"axj\"o University, S-35195, Sweden\\
\emph{email: Andrei.Khrennikov@msi.vxu.se}
\\~\\
\textsf{and}\\~\\
\textsf{Ya.I.~Volovich}
\\
Physics Department, Moscow State University\\
Vorobievi Gori, 119899 Moscow, Russia\\
\emph{email: yaroslav\_v@mail.ru}
}

\date {~}
\maketitle
\thispagestyle{empty}

\begin{abstract}

We consider a classical analogue of the well known  quantum two-slit
experiment. Charged particles are scattered on flat screen with two slits
and hit the second screen. We show that the probability distribution on the
second screen when both slits are open is not given by the sum of 
distributions for each slit separately, but has an extra interference term
that is given with the quantum rule of the addition of probabilistic
alternatives. We show that the proposed classical model has a context
dependence and could be adequately described with contextual formalism.

\end{abstract}

\newpage

\section{Introduction}

It is well known that the classical rule for the addition of probabilistic
alternatives
\begin{equation}
\label{F1}
\P=\P_1+\P_2
\end{equation}
does not work in experiments with elementary particles. Instead of this
rule, we have to use quantum rule
\begin{equation}
\label{F2}
\P=\P_1+\P_2+2\sqrt{\P_1\P_2}\cos\theta.
\end{equation}

The classical rule for the addition of probabilistic alternatives is
perturbed by so called  interference term. The difference between
`classical' and `quantum' rules was (and is) the source of permanent
discussions as well as various misunderstandings, see e.g. on general
references [1]-[19]. We just note that the appearance of the interference
term was the source of the wave-viewpoint to the theory of elementary
particles; at least the notion of superposition of quantum states was
proposed as an attempt to explain the appearance of a new probabilistic
calculus in the two slit experiment, see, for example, Dirac's book [1] on
historical  analysis of the origin of quantum formalism. We also mention
that Feynman interpreted (\ref{F2}) as the evidence of the violation of the
additivity postulate for `quantum probabilities', [5].

In particular, this induced the viewpoint that there are some special 
`quantum' probabilities that differ essentially from ordinary `classical'
probabilities. We also remark that  the orthodox Copenhagen interpretation
of quantum formalism is just an attempt to explain (\ref{F2}) without  to
apply to mysterious `quantum probabilities'. To escape the use of a new
probabilistic calculus, we could  suppose that, e.g. electron participating
in the two slit experiment is in the superposition of passing through both
slits. We mention that, in particular, this implies that quantum particles
do not have trajectories. 

However, there is another approach to quantum experiments that is not so
strongly based on special "non-classical" features of elementary particles.
This is so called contextualist approach. In experiments with elementary
particles we have to take into account whole experimental arrangement, see
N. Bohr [3] and W. Heisenberg [4]. Thus quantum probabilities are
context-depending probabilities. Here the term context is used for a
complex of experimental physical conditions. The contextualist approach to
quantum mechanics was developed in many directions, see e.g. [9] -[19].
Recently the classical probabilistic derivation of  quantum rule (\ref{F2})
was presented in the series of papers of one of the author's [20] -[23].
This derivation demonstrated that it seems that special quantum features
are not  important to get interference modification (\ref{F2}) of
classical rule (\ref{F1}). Interference can be induced for macroscopic
systems  by variations of context. It is not important what kind of
physical systems, micro or macro, are prepared by a complex of physical
conditions. Theoretical investigations [20]-[23] demonstrate that we could,
in principle, get interference for macroscopic systems.

These theoretical considerations stimulated the numerical investigation
presented in this paper. We consider a classical analogue of a well known 
quantum two-slit experiment. Charged particles are scattered on flat screen
with two slits and hit the second screen. We show that the probability
distribution on the second screen when both slits are open is not given by
the sum of  distributions for each slit separately, but has an extra
interference term that is given with the quantum rule of the addition of
probabilistic alternatives. In principle, we can introduce complex
amplitudes of  (classical!) probabilities and work with macroscopic
quantities in the Hilbert space framework, cf. [22].

\section{The Model}

We consider a classical analog of the two slit experiment (Fig.1). The
uniformly charged round particles are emitted at point $e$ with fixed
velocity with the angles evenly distributed in the range $[0,2\pi)$. Each
particle interacts with the uniformly charged flat screen $S_1$. The charge
distribution on the particle and the screen stays unchanged even if the
particle comes close to the screen. Physically this is a good approximation
when the particle and the screen are both made of dielectric. There are two
rectangular slits in the screen (on the Fig.1 the slits are
perpendicular to the plane of the picture). Particles pass through the
slits in screen $S_1$ and gather on screen $S_2$.

\begin{figure}
\begin{center}
\epsfig{file=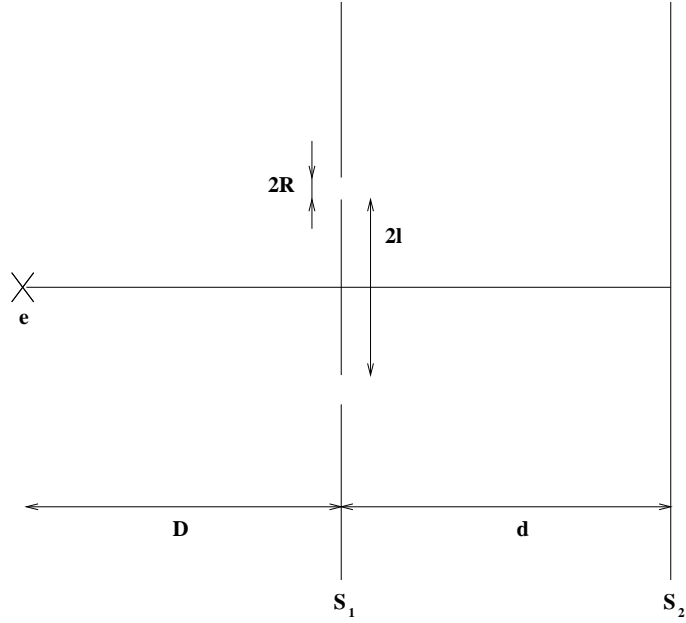,width=9cm}
\end{center}
\caption{Two-slit experiment.}
\end{figure}

We consider three experiments. In the first one the bottom slit is closed
with the shutter, in the second - the upper slit, and in the third both
slits are left open. The charge distribution on the shutter is the same as
on the screen, i.e. in the first two experiments one can think as if the
uniformly charged screen has only one slit. In this and several paragraphs
below by the screen we mean screen $S_1$.

Now let us write the equations of motion in each of three experiments
($i=1,2,3$)
\begin{equation}
\label{eq}
m\ddot{\vect{r}}=\vect{F_i}
\end{equation}
where $\vect{r}$ determines place of the particle. Here $\vect{F_i}$ is
force affecting the particle in each experiment. It is given by the
Coulomb's law
\begin{equation}
\label{ex}
\vect{F_i}=\int\limits_{\mathcal{D}_i}
           \frac{q\sigma}{|\vect{r'}|^2}\cdot\frac{\vect{r'}}{|\vect{r'}|}~
		   ds
\end{equation}
where $\vect{r'}$ is a vector from an element on the screen to the
particle, $q$ is charge of the particle, $\sigma$ is charge density on the
screen, i.e. charge of a unit square. We integrate over the surface of the
screen, the integration region $\mathcal{D}_i$ is plane of the screen
except the splits, as it was mentioned above it is different in each
experiment depending on which slits are opened.

Projecting equations (\ref{eq})-(\ref{ex}) to $xy$-plane, where $x$ and $y$
denotes horizontal and vertical coordinates of the particle respectively we
get
\begin{equation}
\label{em}
\begin{split}
m\ddot{x} &= q\sigma \int\limits_{\Gamma_i} dy' \int\limits_{\mathbb{R}}
dz' \frac{x}{(x^2+(y-y')^2+{z'}^2)^{3/2}}\\
m\ddot{y} &= q\sigma \int\limits_{\Gamma_i} dy' \int\limits_{\mathbb{R}}
dz' \frac{y-y'}{(x^2+(y-y')^2+{z'}^2)^{3/2}}
\end{split}
\end{equation}
where $\Gamma_i$ indicates the integration region for the $i$-th
experiment. In our previous notations
$\mathcal{D}_i=\Gamma_i\times\mathbb{R}$. We have 
\begin{equation}
\label{eg}
\begin{split}
\Gamma_1&=(-\infty,l)\cup(l+2R,+\infty)\\
\Gamma_2&=(-\infty,-l-2R)\cup(-l,+\infty)\\
\Gamma_3&=(-\infty,-l-2R)\cup(-l,l)\cup(l+2R,+\infty)
\end{split}
\end{equation}
Here $2l$ is the distance between slits and $2R$ is the height of the slit.

Integrating the rhs of (\ref{em}) we get
\begin{equation}
\label{ew}
\begin{split}
m\ddot{x} &= q\sigma \sum_{(a,b)\subset\Gamma_i}
     2\left( \arctan{\frac{b-y}{x}}-\arctan{\frac{a-y}{x}} \right)\\
m\ddot{y} &= q\sigma \sum_{(a,b)\subset\Gamma_i}
     \ln\left(x^2+(b-y)^2\right)-\ln\left(x^2+(a-y)^2\right)
\end{split}
\end{equation}
where the notation $(a,b)\subset\Gamma_i$ means that the sum extends over
all subranges of $\Gamma_i$ given in (\ref{eg}). For example for $i=1$ we
have two summands with $(a=-\infty,b=l)$ and $(a=l+2R,b=+\infty)$, and
(\ref{ew}) will take the following form
\begin{equation}
\label{ez}
\begin{split}
m\ddot{x} &= 2q\sigma
     \left(\pi+\arctan{\frac{l-y}{x}}-\arctan{\frac{l+2R-y}{x}}\right)\\
m\ddot{y} &= q\sigma
     \ln\frac{x^2+(l-y)^2}{x^2+(l+2R-y)^2}
\end{split}
\end{equation}
here we took into account that $\arctan(\pm\infty)=\pm\pi/2$ and the sum of
the logarithms for $a=-\infty$ and $b=+\infty$ vanishes.

We take the following initial values
\begin{equation}
\label{ic}
\begin{aligned}
x(0) &=-D\\
y(0) &= 0
\end{aligned}
\qquad
\begin{aligned}
\dot{x}(0) = v_0\cos\alpha\\
\dot{y}(0) = v_0\sin\alpha
\end{aligned}
\end{equation}
where angle $\alpha$ is a random variable uniformly distributed in
$[0,2\pi)$. The constant parameters $v_0$ and $D$ are initial velocity and
distance between emitter and the screen.

Particles are emitted at point $e$ (see Fig.1), move obeying
(\ref{ew}),(\ref{ic}) passing through slit(s) in the screen $S_1$ and
gather on the screen $S_2$. Having points where particles hit the screen
$S_2$ we compute frequencies with which particles appear on screen $S_2$ as
a function of coordinates on the screen. We interpret this frequencies as
probability distributions. We are interested in computing the probability
distribution over a vertical line on screen $S_2$ with $z=0$. That is why
we consider a motion only in the $xy$-plane and initial values (\ref{ic})
do not contain $z$-coordinate.

We solve the equations of motion (\ref{ew}) with initial conditions
(\ref{ic}) numerically. We use Rungie-Kutta 4th order switching to Adams 4th
order method. We used GNU C++ (g++) compiler to realize the simulation on
Ultra-SPARC computer running Solaris. We had to explore about $10^5$
trajectories and we used 4-processor parallel computer located an V\"axj\"o
University. The computation process was easy to make parallel as moving
particles are not interacting, i.e. one could think as if they were emitted
with long intervals. The algorithm automatically adjusted the computation
precision making shorter steps when the particle comes near to the first
screen or the coordinates ($x$ and $y$) changed more than minimum precision
allowed. The first stage of computation was calibration when the algorithm
determined the angle ranges for which the particles passed through the
slits and hit the second screen. This reduced the angle range from
$[0,2\pi)$ to a set of ranges, which are different in each experiment. In
fact we used symmetry of the first two ones (when only upper or lower slit
is opened) making computations only for the first one. The second screen
was separated with cells of equal size, the diameter of a particle. The
number of particles which hit into each cell was calculated and interpreted
as a probability distribution.

\begin{figure}
\begin{center}
\epsfig{file=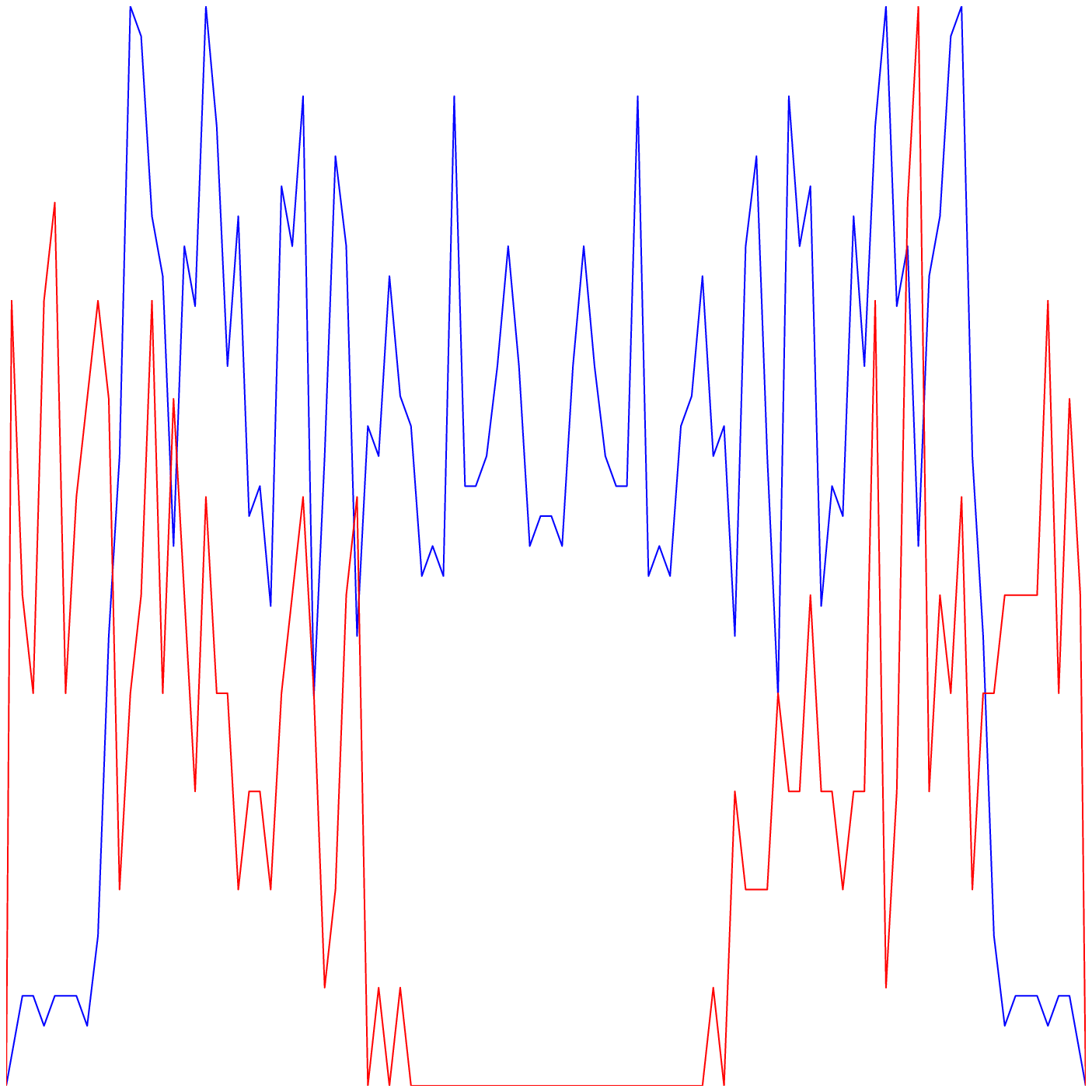,width=9cm,height=2cm}
\end{center}
\caption{Probability distributions. Blue line is
$\left(\frac{\displaystyle P_1}{\displaystyle2}+
       \frac{\displaystyle P_2}{\displaystyle 2}\right)$, red line
is $P_{12}$.}
\end{figure}

Let us denote the probability distribution in the first experiment (only
upper slit is opened) as $P_1=P_1(y)$, in the the second experiment (only
lower slit is opened) as $P_2=P_2(y)$, and in the third experiment (both
slits are opened) as $P_{12}=P_{12}(y)$. Although since the force is
different in each experiment, see (\ref{eq}), it is quite clear that
(Fig.2)
\begin{equation}
P_{12}\neq\frac{P_1}{2}+\frac{P_2}{2}
\end{equation}
To become an equality the above equation should have an extra term
\begin{equation}
\label{it}
P_{12}=\frac{P_1}{2}+\frac{P_2}{2}+\sqrt{P_1 P_2}\cos\theta
\end{equation}
where $\sqrt{P_1 P_2}\cos\theta$ is a so-called interference term (Fig.3),
and $\theta=\theta(y)$ is spread along $y$-axis.

\begin{figure}
\begin{center}
\epsfig{file=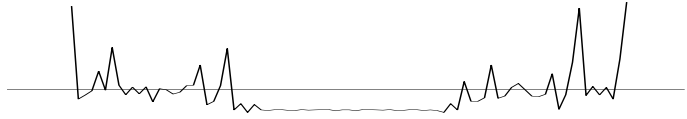,width=9cm,height=2.9cm}
\end{center}
\caption{Interference term,
$\cos\theta=(2P_{12}-(P_1+P_2))/\sqrt{P_1P_2}$.}
\end{figure}

The function
\begin{equation}
\cos\theta=\frac{2P_{12}-(P_1+P_2)}{\sqrt{P_1P_2}}
\end{equation}
is shown on the (Fig.2). Please note that as there are ranges where $P_1$
or $P_2$ are equal to zero, i.e. $P_1P_2=0$ the function $\cos\theta$ is
not determined and from (\ref{it}) we see that $P_{12}$ does not depend on
it.

{\bf Conclusion.} We have shown that the proposed classical  model has a
context dependence and could be adequately described with contextual
formalism. {\it Quantum like behavior for macro systems is demonstrated.}
We simulate  quantum-like interference for macroscopic objects. Such a
simulation essentially  reduced the gap between micro and macro worlds.
\footnote{In particular,recent experiments of the group of A. Zeilinger
[23] and the Boulder-group [24] can be interpreted as successful steps in
this direction.} 

One of the authors (A.K.) would like to thank S. Albeverio, L. Accardy,  L.
Ballentine, V. Belavkin, E. Beltrametti, G. Cassinelli, A. Chebotarev, W.
De Baere, W. De Myunck, R. Gill, D. Greenberger, S. Goldstein, C. Fuchs, L.
Hardy, A. Holevo, T. Hida, P. Lahti, E. Loubents, D. Mermin, T. Maudlin, 
A. Peres, I. Pitowsky, A. Plotnitsky, A. Shiryaev, O. Smoljanov, J.
Summhammer, L. Vaidman and I. Volovich,  A. Zeilinger for fruitful
discussions on  probabilistic foundations of quantum mechanics.

This work was done during the visit of Ya.V. to V\"axj\"o University, he is
grateful for the kind hospitality.

{\bf References}

[1] P. A. M.  Dirac, {\it The Principles of Quantum Mechanics}
(Oxford Univ. Press, 1930).

[2] W. Heisenberg, {\it Physical principles of quantum theory.}
(Chicago Univ. Press, 1930).

[3]  N. Bohr, {\it Phys. Rev.,} {\bf 48}, 696-702 (1935).

[4] J. von Neumann, {\it Mathematical foundations
of quantum mechanics} (Princeton Univ. Press, Princeton, N.J., 1955).

[5] R. Feynman and A. Hibbs, {\it Quantum Mechanics and Path Integrals}
(McGraw-Hill, New-York, 1965).

[6] J. M. Jauch, {\it Foundations of Quantum Mechanics} (Addison-Wesley,
Reading, Mass., 1968).

[7]  P. Busch, M. Grabowski, P. Lahti, {\it Operational Quantum Physics}
(Springer Verlag, 1995).

[8] B. d'Espagnat, {\em Veiled Reality. An anlysis of present-day
quantum mechanical concepts} (Addison-Wesley, 1995). 

[9] A. Peres, {\em Quantum Theory: Concepts and Methods} (Kluwer Academic
Publishers, 1994).

[10] E. Beltrametti  and G. Cassinelli, {\it The logic of Quantum mechanics.}
(Addison-Wesley, Reading, Mass., 1981).

[11]  L. E. Ballentine, {\it Quantum mechanics} (Englewood Cliffs, 
New Jersey, 1989).

[12] A.Yu. Khrennikov, {\it Interpretations of 
probability} (VSP Int. Publ., Utrecht, 1999).

[13] S. P. Gudder,  J. Math Phys., {\bf 25}, 2397 (1984); S. P. Gudder, N.
Zanghi,
 Nuovo Cimento B {\bf 79}, 291(1984).

[14] L. Accardi, 
Phys. Rep., {\bf 77}, 169(1981); L. Accardi,
in {\it   Stochastic processes in quantum theory and statistical physics,}
edited by  S. Albeverio et al., Springer LNP {\bf 173} 1 (1982).
L. Accardi, {\it Urne e Camaleoni: Dialogo sulla realta,
le leggi del caso e la teoria quantistica.} Il Saggiatore, Rome (1997).
L.Accardi, P. Regoli: Quantum probability
versus non-locality: crucial experiment. Preprint, Centro V. Volterra,
N. 409, 2000.

[15] I. Pitowsky,  Phys. Rev. Lett, {\bf 48}, N.10, 1299(1982).

[16]  A. Fine,  Phys. Rev. Letters, {\bf 48}, 291 (1982);
P. Rastal, Found. Phys., {\bf 13}, 555 (1983).

[17] W. De Baere,  Lett. Nuovo Cimento, {\bf 39}, 234 (1984);
{\bf 25}, 2397(1984); 
 W. De Muynck, W. De Baere, H. Martens,
 Found. of Physics, {\bf 24}, 1589 (1994);

[18] L. Ballentine, Probability theory in quantum mechanics. {\it American
J. of Physics,} {\bf 54}, 883-888 (1986).

[19] J. Summhammer, Int. J. Theor. Physics, {\bf 33}, 171 (1994);
Found. Phys. Lett. {\bf 1}, 113 (1988); Phys.Lett., {\bf A136,} 183 (1989).

[20] A. Yu. Khrennikov, {\it Ensemble fluctuations and the origin of
quantum probabilistic
rule.} Rep. MSI, V\"axj\"o Univ., {\bf 90}, October (2000).

[21] A. Yu. Khrennikov, {\it Classification of transformations of
probabilities for preparation procedures:
trigonometric and hyperbolic behaviours.} Preprint quant-ph/0012141, 24 Dec
2000.

[22] A. Yu. Khrennikov, {\it Linear representations of probabilistic
transformations induced
by context transitions.} Preprint quant-ph/0105059, 13 May 2001.

[23] A. Yu. Khrennikov, {\it `Quantum probabilities'  as context depending
probabilities.}
Preprint quant-ph/0106073, 13 June 2001.

[24] A. Zeilinger, {\it Recent results in fullerene Inteferometry and in
quantum teleportation.}
Abstracts of Int. Conf. {\it Exploring Quantum Physics}, Venice-2001.

[25] A. Ben-Kish, J. Britton, D. Kielpinski, D. Leibfried,
V. Meyer, M. Rowe, C. Sakket, W. Itano, C. Monroe, D. Wineland,
{\it Ion entanglement experiments at NIST-Boulder.}
Abstracts of Int. Conf. {\it Exploring Quantum Physics}, Venice-2001.

\end{document}